# A Comparative Analysis of Nigeria's Power Sector with and without Grid-Scale Storage: Future Implications for Emission and Renewable Energy Integration

Stanley Aimhanesi Eshiemogie [a,*], Peace Precious Aielumoh [b], Tobechukwu Okamkpa [c], Miracle Chinonso Jude [a], Lois Efe [d], Andrew Nosakhare Amenaghawon [a,*]

[a] *Department of Chemical Engineering, University of Benin, Benin City, Edo State, Nigeria*

[b] *Department of Computer Engineering, University of Benin, Benin City, Edo State, Nigeria*

[c] *Department of Mechanical Engineering, University of Nigeria, Nsukka, Enugu State, Nigeria*

[d] *Department of Electrical and Electronics Engineering, University of Manchester, Manchester M13 9PL, UK*

**Abstract**

This work compares two electricity scenarios for Nigeria by 2050, focusing on the inclusion and exclusion of electricity storage technologies, using a machine learning-supported approach. A Central Composite Design (CCD) was used to generate a design matrix for data collection, with EnergyPLAN software used to create energy system simulations on the CCD data for four outputs: total annual cost, $CO_2$ emissions, critical excess electricity production (CEEP), and electricity import. Three machine learning (ML) algorithms—support vector regression (SVR), extreme gradient boosting (XGBoost), and multi-layer perceptron (MLP)—were tuned using Bayesian optimization to develop models mapping the inputs to outputs. A genetic algorithm was employed for multi-objective optimization to determine the optimal input capacities that minimize the outputs. Results indicated that incorporating electricity storage technologies (EST) leads to a 37% increase in renewable electricity sources (RES) share, resulting in a 19.14% reduction in $CO_2$ emissions. EST such as battery energy storage systems (BESS), pumped hydro storage (PHS), and vehicle-to-grid (V2G) storage allow for the storage of the critical excess electricity that comes with increasing RES share. Integrating EST in Nigeria's 2050 energy landscape is crucial for incorporating more renewable electricity sources into the energy system – thereby reducing $CO_2$ emissions – and managing excess electricity production. This study outlines a plan for optimal electricity production to meet Nigeria’s 2050 demand, highlighting the need for a balanced approach that combines fossil fuels, renewable energy, nuclear power, and advanced storage solutions to achieve a sustainable and efficient electricity system.

[*]stanley.eshiemogie@eng.uniben.edu (corresponding author)

[*]andrew.amenaghawon@uniben.edu (corresponding author)

## Graphical abstract

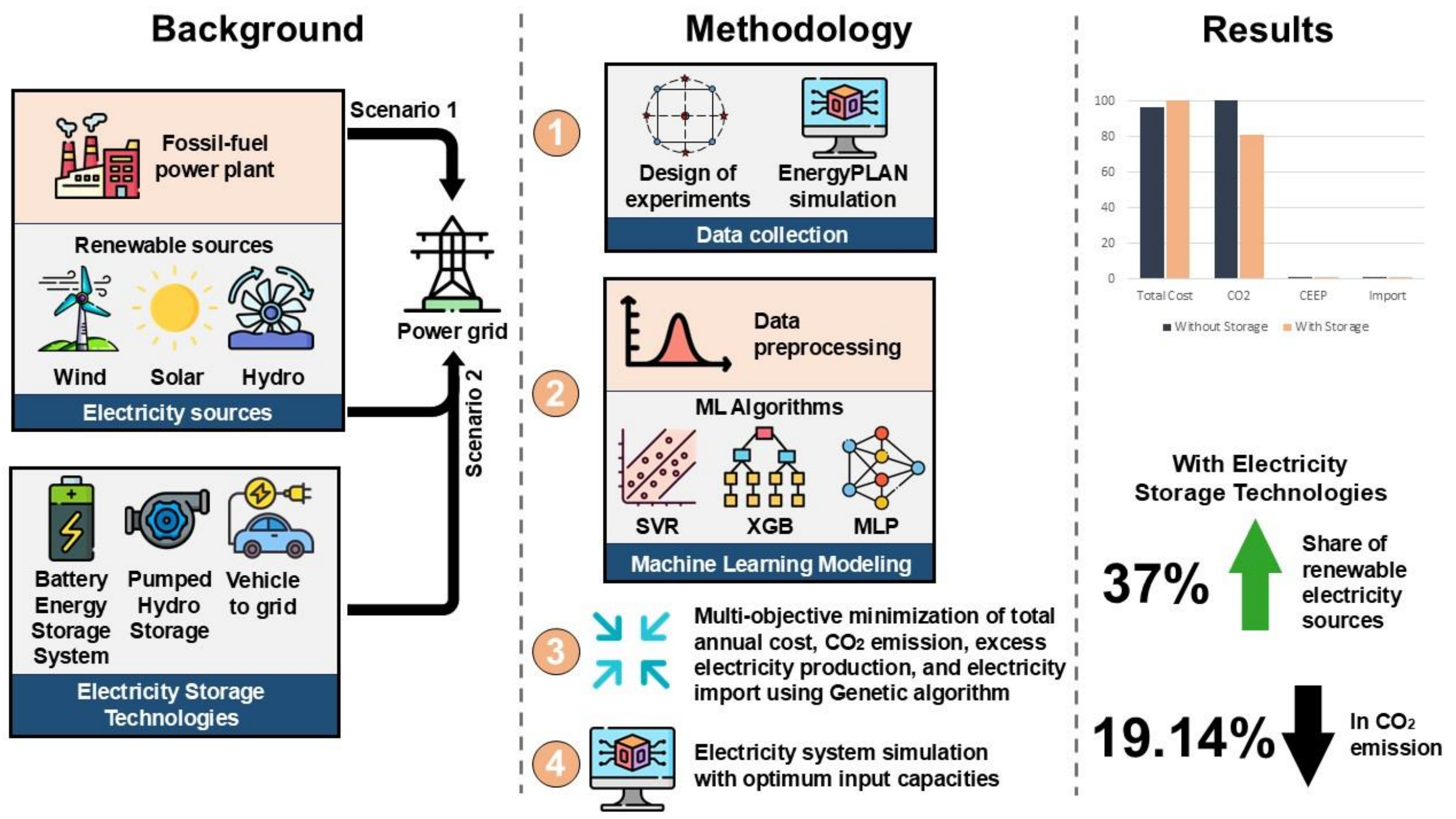

Keywords: Electricity storage, machine learning, meta-heuristic optimization, renewable electricity sources

## 1. Introduction

Nigeria’s current electricity infrastructure is plagued by significant challenges, with over 60% of the population lacking access to the grid, and even those connected experience frequent power outages due to insufficient generation capacity (Bello et al., 2023). Despite boasting an installed total generation capacity of 14.38 GW, only 6 GW is available for use (Chanchangi et al., 2023). This shortfall creates a scenario where there is plenty of potential capacity, but the reality is one of scarcity. Various factors contribute to this gap, including aging infrastructure, maintenance challenges, and gas supply constraints for thermal power plants. Only a fraction of the electricity generated reaches the end-user, with a significant portion, amounting to over 50%, lost during the generation and transmission phases (Adamu et al., 2022; Komolafe & Udofia, 2020; Oladunni Juliet, 2019).

Nigeria currently relies on two main sources for its large-scale electricity generation: natural gas and hydro (Abe et al., 2024). With petroleum resources accounting for as much as 70% of the country's commercial primary energy (Aliyu et al., 2013), this dependence burdens the environment with high carbon emissions and accelerates climate change. Furthermore, the country is currently unable to meet the electricity demand of 31.2 GW, with a supply of 6 GW (Chanchangi et al., 2023), leaving over 60% of the population without electricity, and regular load shedding among the population with electricity. Moreover, this demand has been estimated to rise in the coming years, reaching over 900 TWh in 2050 (Electricity Commission of Nigeria, 2015).

Renewable electricity sources (RES) like hydro, solar and wind offer a way forward. These sources have been shown to provide the electricity needed to meet demand in some countries like the USA, UK, Canada, and Japan (Tansel Tugcu & Menegaki, 2024), and have been employed for over 70% electrification of Iceland (International Energy Agency, 2022a) and 90% of Norway (International Energy Agency, 2022b). In the global drive towards decarbonization, RES have been shown to be very effective alternatives to fossil fuels. The contribution of RES to the national grid in the Nigerian electricity landscape remains significantly low (Tambari et al., 2020), with hydropower being the only RES with a substantial contribution to the grid (Owebor et al., 2021). Recognizing this limitation, Nigeria aims to bridge the generation gap and promote economic growth by further developing hydropower and other RES like wind and solar power. However, the integration of RES solutions faces the challenge of variable supply, as there is no 24-hour sunlight, and windspeed varies with time of the year and location. This variability leads to critical excess electricity production (CEEP), during times of high renewable energy generation and low demand. RES integration in Nigeria faces several other challenges, including a lack of expertise, uncertainty of financial resources, poor commitment, and a lack of government investment in these technologies (Oyedepo et al., 2018). Overcoming these challenges is essential for Nigeria to fully realize the potential of clean energy and achieve the 2050 climate vision to reduce the current emission by 50% (Department of Climate Change, 2021).

Through investment in grid-scale electricity storage, Nigeria can enhance the flexibility of its grid to accommodate a higher proportion of RES, thus contributing to a more sustainable electricity supply (Matthew et al., 2018a). To achieve a sustainable electricity future, future power plants need to be adaptable, generating electricity when demand is high, and with GSES also releasing electricity when demand is high and efficiently storing excess electricity from variable renewable electricity sources (VRES) during periods of low demand (Baigorri et al., 2023), or high VRES supply (Bamisile et al., 2020). According to Matthew et al., (2018), optimizing the contribution of renewable energy to a sustainable electricity future depends on effectively integrating these sources with complementary storage technologies. Furthermore, storage empowers self-consumption and distributed generation, enabling households and businesses to utilize their own renewable energy production for on-site needs and potentially feed excess energy back into the grid (Dioha et al., 2022).

Electricity storage technologies (EST) such as battery energy storage systems (BESS), pumped hydro storage (PHS) and vehicle-to-grid (V2G) storage offer significant potential as clean energy storage solutions for electric power systems, particularly in scenarios with a high mix of VRES like wind and solar power (M. O. Dioha et al., 2022). It also enhances grid stability, regulates voltage, and provides long discharge durations, reducing wasted energy, congestion, and emissions (Mahfoud et al., 2023). The successful implementation of energy storage solutions, alongside advanced system management, can overcome the challenges associated with VRES (Oyedepo et al., 2018).

Due to the high complexity of energy systems, having different generating sources and storage technologies, energy planning today requires a comprehensive approach that combines economic and engineering insights. Research highlights that power generation models are now essential tools in this field. Power planning models are utilized globally for energy system analysis and optimization. EnergyPLAN software stands out for its versatility, ability to model 100% renewable energy-systems, user-friendliness, and comprehensive approach of optimizing system operation (Connolly et al., 2010). It has been employed to analyse combined heat and power (CHP) production for regions globally (Østergaard et al., 2022). Prina et al. (2020) used the EnergyPLAN software to model the optimal future energy mix in Niederösterreich, aiming for an 80% reduction in $CO_2$ emissions. They explored scenarios with and without electric mobility, finding that higher electric vehicle (EV) penetration significantly increased electricity demand but reduced overall system costs by approximately 10% due to the efficiency of electric motors and reduced fossil fuel use. Additionally, EnergyPLAN was instrumental in analysing various scenarios, including dry and wet years, to understand their impact on Portugal's electricity generation mix. Results showed that renewable sources could meet up to 89% and 100% of demand, respectively, with hydropower and interconnections playing a key role. The study also indicated that this high share of renewables could reduce carbon intensity by 97% (Graça Gomes et al., 2020). These approaches have been applied to various power system applications, such as load forecasting, fault detection, and condition monitoring (Strielkowski et al., 2023). However, very few works have applied ML to EnergyPLAN modelling (Punyam Rajendran & Gebremedhin, 2024; Yang et al., 2021). Moreover, to the best of our knowledge, no work has focused on applying Bayesian optimization for tuning the hyperparameters of ML models used in EnergyPLAN modelling. Furthermore, the integration of ML modelling with other energy simulation tools such as EnergyPLAN for modelling future electricity scenarios has not been applied in the Nigerian context.

Several studies have incorporated modelling tools to explore potential pathways for Nigeria's energy transition. Oyewo et al. (2018) investigated a cost-optimized pathway for a complete transition to 100% renewable energy by 2050 using the LUT energy system model. They simulated different policy scenarios, finding that best policy scenarios (BPS) focusing on renewables lead to initially higher levelized costs of electricity (LCOE), but these costs are expected to drop significantly by 2050 to 46-48 €/MWh. In contrast, current policy scenarios (CPS) relying on fossil fuels had lower initial costs but was simulated to result in higher long-term LCOE. However, this study does not assess storage capacities and how they affect the overall energy system. Another study employing The Integrated MARKAL-EFOM System (TIMES) model reinforces the feasibility of a full transition to renewables. This research explores two scenarios: the Business-As-Usual (BAU), which acts as a benchmark for comparing alternative policy options, and scenarios aiming for 100% Renewable Energy (REN) electricity generation in Nigeria. By 2050, installed capacities in BAU scenarios peak at 411.2 GW, primarily driven by solar PV and gas turbines. In contrast, REN scenarios depict substantial increases, reaching 1029.9 GW, with solar PV and wind turbines dominating the landscape (Tambari et al., 2020). Bamisile et al. (2020) used the EnergyPLAN tool to design a sustainable pathway for the 100% electrification of Nigeria by 2030, with a total annual cost of 19.2 BUSD and an annual $CO_2$ emission of 77.27 Mt. for a demand of 200 TWh. Their approach, however, did not take into consideration the optimization of all outputs simultaneously.

While many researchers have worked with advanced modelling tools to model Nigeria’s future electricity landscape, no work has assessed the 2050 electricity landscape and compared the scenarios with and without electricity storage by using a meta-heuristic algorithm for the multi-objective optimization of total annual cost, $CO_2$ emission, CEEP, and electricity import. This analysis is crucial as it identifies areas needing investment and policy intervention, guiding sustainable energy decisions for Nigeria. It addresses challenges in renewable energy variability and intermittency for grid-scale demand. A comprehensive analysis of both scenarios is essential to explore trade-offs and optimal pathways for resource efficiency and societal benefits. This study lays the groundwork for further research in technology development and informed strategic decision-making for Nigeria's sustainable energy future. Furthermore, no work has followed an approach employing data generation through experimental design, simulation using the EnergyPLAN software, modelling using ML tools, and multi-objective optimization using meta-heuristic algorithm. This forms the major thrust of the novelty of this work.

As Nigeria aspires to achieve sustainable development goals and enhance energy security, there is a pressing need for a comprehensive assessment of the country's electricity storage capacity. This study employs a comprehensive approach combining design of experiments, ML modelling, energy systems simulation, and multi-objective optimization to compare two electricity scenarios for Nigeria in 2050: with electricity storage technologies (WEST) and without electricity storage technologies (WoEST). This work evaluated several ML algorithms, including SVR, XGBoost, and MLP, with the hyperparameters tuned using Bayesian optimization. Subsequently, multi-objective optimization was employed for minimizing total annual costs, $CO_2$ emissions, CEEP, and electricity imports for both scenarios to facilitate a robust comparison. Additionally, the importance and contributions of input features to model performance were assessed using Shapley values, a corporative game theory concept used to fairly distribute the total gains among the players based on their individual contributions. By leveraging predictive modelling and state-of-the-art analytics, this study analyses the requirements and devises effective strategies tailored to Nigeria's unique energy landscape, promoting cost-effective integration of renewable energy sources with grid-scale storage solutions for Nigeria by 2050.

## 2. Methodology

### 2.1. Framework for electricity scenario analysis

A systematic approach was employed in this study for analyzing two electricity scenarios for Nigeria by 2050. The first electricity scenario, WoEST, was characterized by the use of natural gas power plants (PP), nuclear power plants, and RES including wind, river hydro, photo-voltaic, and concentrated solar power, while the second electricity scenario, WEST, was characterized by the aforementioned electricity sources, in addition to storage technologies such as BESS, PHS, and V2G. Both scenarios were first simulated using the EnergyPLAN software. As presented in Fig. 1, the modeling followed a sequential pattern, from experimental design, to in-silico experimentation for data collection, to machine learning modeling, to multi-objective optimization to determine the most optimal capacities for the various technologies that minimize total annual cost, $CO_2$ emission, CEEP and electricity import, and to analysis to investigate the effects of different technologies on the outputs. Total annual cost combines variable costs, fixed operation costs, and annual investment costs, providing a comprehensive measure of economic feasibility. $CO_2$ emissions were examined to assess the environmental impact of a system, ensuring that the proposed systems align with climate goals. The efficiency of each system in managing excess power generation was measured through CEEP, which accounts for the excess electricity produced that exceeds demand and transmission line capacity, highlighting areas for potential waste reduction. Additionally, the reliance on other countries for electricity was evaluated through electricity import, which has crucial implications for energy security and self-sufficiency. These metrics were chosen for their critical importance in evaluating the performance and sustainability of the electricity system.

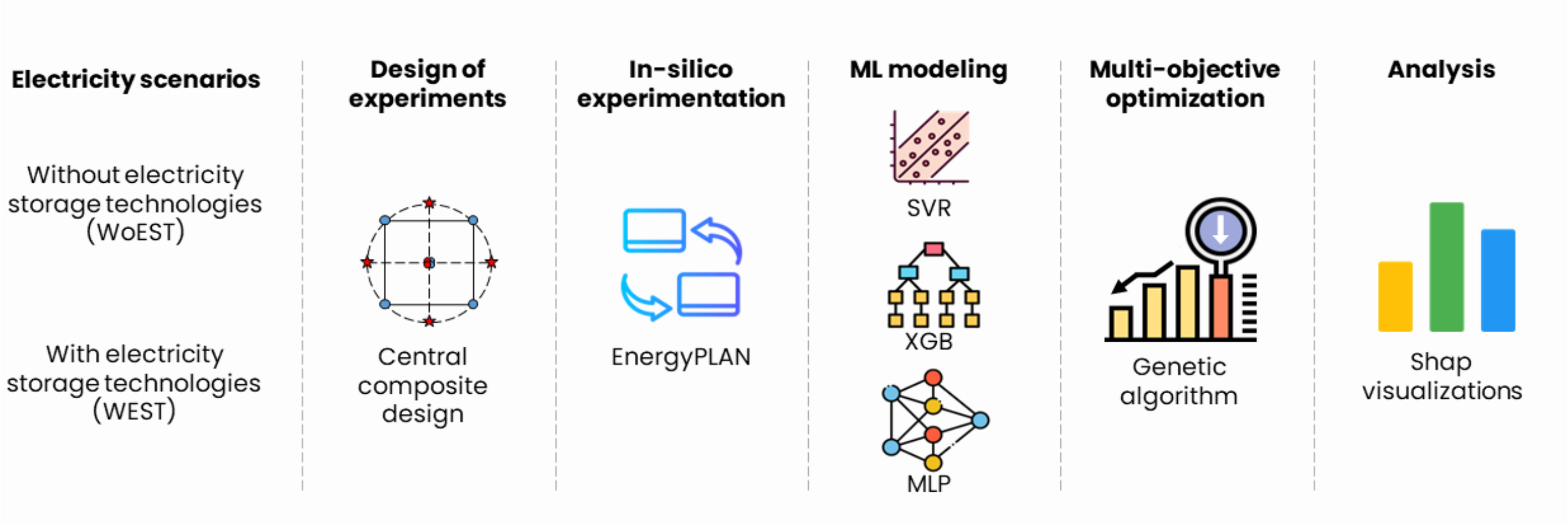

Fig. 1: Research workflow for the study scenarios

### 2.2. Data generation

Design of experiments (DOE) was used to generate the data that was later used to simulate and model the electricity scenarios. DOE is a systematic method used to determine the relationship between factors affecting a process and the output of that process (Politis et al., 2017). Specifically, a central composite design (CCD) implemented using the pyDOE library in the Python programming language was applied to plan the in-silico experiments for data collection. The CCD is made up of the factorial points, which are possible combinations of points between pre-specified high and low levels; the axial points, which are determined by the rotatability, a property that allows constant variance prediction at all points that are equidistant from the design center, and represented by alpha (Malekzadeh & Fatemi, 2015); and the center points, which are used to generate quadratic model terms and to analyze the quadratic interaction effect between the factors (Szpisják-Gulyás et al., 2023). The total number of experimental runs, $N$, generated by a CCD design matrix is given by Equation (1), where $k$ is the number of factors, $2^k$ is the factorial term, $2k$ is the axial term, and $n_c$ is the center points term (Tyagi & Kumar, 2024).

$$N = 2^k + 2k + n_c \tag{1}$$

CCD is a more detailed and accurate experimental approach that includes factorial and center point experiments alongside axial points. This design provides more precise predictions for the optimization of input variables, making it a more preferred technique compared to other response designs (Tyagi & Kumar, 2024).

In the WoEST scenario, the energy mix included fossil power plants (PP) alongside renewable sources such as nuclear, wind, photovoltaic (PV), river hydro (RH), and concentrated solar power (CSP). In addition to the technologies in the WoEST scenario, the WEST scenario featured electric vehicle (EV) storage, pumped-hydro storage (PHS), and battery energy storage systems (BESS). EV storage was defined by total vehicle battery storage (VBS) capacity, grid-to-vehicle (G2V) capacity for vehicle charging, and vehicle-to-grid (V2G) capacity for grid supply. PHS was characterized by PHS storage capacity (PHS SC), PHS charge capacity (PHS CC), and PHS discharge capacity (PHS DC). Similarly, BESS included BESS storage capacity (BESS SC), BESS charge capacity (BESS CC), and BESS discharge capacity (BESS DC).

### 2.3. Experimental design for electricity scenarios

The CCD was employed as the first step in a series of steps towards identifying the optimal capacities for renewable and non-renewable electricity sources, as well as storage technologies that minimize total annual cost, $CO_2$ emission, CEEP, and electricity import for each scenario. For the WoEST scenario, a full type factorial core with 10 center points was chosen for the design, making a total of 86 experimental runs with high and low levels matching the ranges of each independent factor as depicted in Table 1. The high levels correspond to the estimated available capacities of each technology in Nigeria by 2050 (Electricity Commission of Nigeria, 2015).

Table 1: CCD configuration for the WoEST scenario

| Name | Units | Low | Middle | High | -0.1 | +0.1 |
|---|---|---|---|---|---|---|
| PP | MW | 50000 | 100,000 | 150000 | 95000 | 105000 |
| Nuclear | MW | 0 | 15,000 | 30000 | 13500 | 16500 |
| Wind | MW | 0 | 7,500 | 15000 | 6750 | 8250 |
| PV | MW | 0 | 100,000 | 200000 | 90000 | 110000 |
| RH | MW | 0 | 10,000 | 20000 | 9000 | 11000 |
| CSP | MW | 0 | 25,000 | 50000 | 22500 | 27500 |

For the experiment with storage technologies, a 1/64 fraction type factorial core with 10 center points was chosen, which resulted in 556 experimental runs, with the ranges of the independent variables matching the high and low levels as depicted in Table 2.

Table 2: CCD configuration for the WEST scenario

| Variables | Units | Low | Middle | High | -0.1 | +0.1 |
|---|---|---|---|---|---|---|
| PP | MW | 50000 | 100000 | 150000 | 95000 | 105000 |
| Nuclear | MW | 0 | 15000 | 30000 | 13500 | 16500 |
| Wind | MW | 0 | 7500 | 15000 | 6750 | 8250 |
| PV | MW | 0 | 100000 | 200000 | 90000 | 110000 |
| RH | MW | 0 | 10000 | 20000 | 9000 | 11000 |
| CSP | MW | 0 | 25000 | 50000 | 22500 | 27500 |
| G2V | MW | 5700 | 27850 | 50000 | 25635 | 30065 |
| VBS | GWh | 10 | 30 | 50 | 28 | 32 |
| V2G | MW | 5000 | 27500 | 50000 | 25250 | 29750 |
| PHS CC | MW | 30000 | 65000 | 100000 | 61500 | 68500 |
| PHS DC | MW | 30000 | 65000 | 100000 | 61500 | 68500 |
| PHS SC | GWh | 0 | 5000 | 10000 | 4500 | 5500 |
| BESS CC | MW | 30000 | 65000 | 100000 | 61500 | 68500 |
| BESS DC | MW | 30000 | 65000 | 100000 | 61500 | 68500 |
| BESS SC | GWh | 0 | 5000 | 10000 | 4500 | 5500 |

**2.4. Energy system modeling**

In-silico experimentation for both electricity scenarios were conducted using EnergyPLAN software (version 16.3), a sophisticated input/output simulation tool designed for comprehensive energy systems analysis (Lund et al., 2021a). This tool facilitates high-resolution temporal analysis, conducting simulations in hourly increments over an entire year, thereby enabling a detailed examination of energy system dynamics (Bamisile et al., 2020). The analysis considers various technical and market-economic simulation strategies, providing insights into the impacts of different energy policies and configurations.

EnergyPLAN requires a diverse array of input parameters, including demands, plant capacities, RES capacities, fuel types, associated costs, and several optional simulation strategies. Key outputs from the software encompass energy balances, annual production metrics, total annual costs, $CO_2$ emissions, fuel consumption, and electricity import/export (Bamisile et al., 2020). While EnergyPLAN is adept at analyzing existing energy systems, its primary application lies in modeling future energy systems with a significant emphasis on the integration of RES. It has been extensively utilized in numerous studies to model 100% renewable energy systems, conduct technical analyses, perform market exchange evaluations, and undertake feasibility studies (Icaza et al., 2021). For this study, fixed inputs included various demand parameters such as energy demand, cooling demand, and transportation demand, as well as fixed costs, operation and maintenance costs, fuel costs, and systems' lifetimes, as detailed in Table 3 and Table 4. According to (Electricity Commission of Nigeria, 2015), electricity demand is projected to be 946 TWh by 2050, which includes a cooling demand of 137 TWh and a transportation demand of 50 TWh, as detailed in Fig. 2 on an hourly time step.

Table 3: Economic description of electricity sources and storage technologies (EnergyPLAN, 2018)

| Type | Units | Investment Cost (M$/unit) | Fixed O&M (% of investment) | Period (Years) |
|---|---|---|---|---|
| PP | MW | 0.99 | 3.05 | 20 |
| Nuclear | MW | 3.595 | 3.5 | 20 |
| Wind | MW | 1.32 | 2.97 | 20 |
| PV | MW | 1.3 | 0.6 | 20 |
| RH | MW | 3.3 | 2 | 50 |
| CSP | MW | 5.98 | 7.7 | 20 |
| PHS charge | MW | 0.6 | 1.5 | 50 |
| PHS discharge | MW | 0.6 | 1.5 | 50 |
| PHS storage | GWh | 7.5 | 0.63 | 50 |
| BESS storage | GWh | 11.19 | 1 | 15 |

Table 4: Fuel cost (EnergyPLAN, 2018)

| Fuel | Price (USD/GJ) | Handling cost (USD/GJ) | |
|---|---|---|---|
| | | To power station | To Biomass conversion plant |
| Coal | 3.1 | 0.26 | N/A |
| Natural gas | 9.1 | 0.41 | N/A |
| Biomass | 6.2 | 1.19 | 1.19 |
| Nuclear | 1.75 | N/A | N/A |

Furthermore, the types of fuels used for electricity generation in Nigeria, as well as the available portfolio, which is expected to include both renewable and non-renewable sources, such as natural gas plants, nuclear power, wind turbines, photovoltaic systems, river hydroelectric power, and concentrated solar power, were gathered from the ECN 2050 forecast (Electricity Commission of Nigeria, 2015). The variable inputs which included the capacities of the renewable and non-renewable electricity sources, as well as energy storage technologies, were derived from the central composite design matrix, which facilitated the exploration of multiple scenarios and their impacts on the energy system.

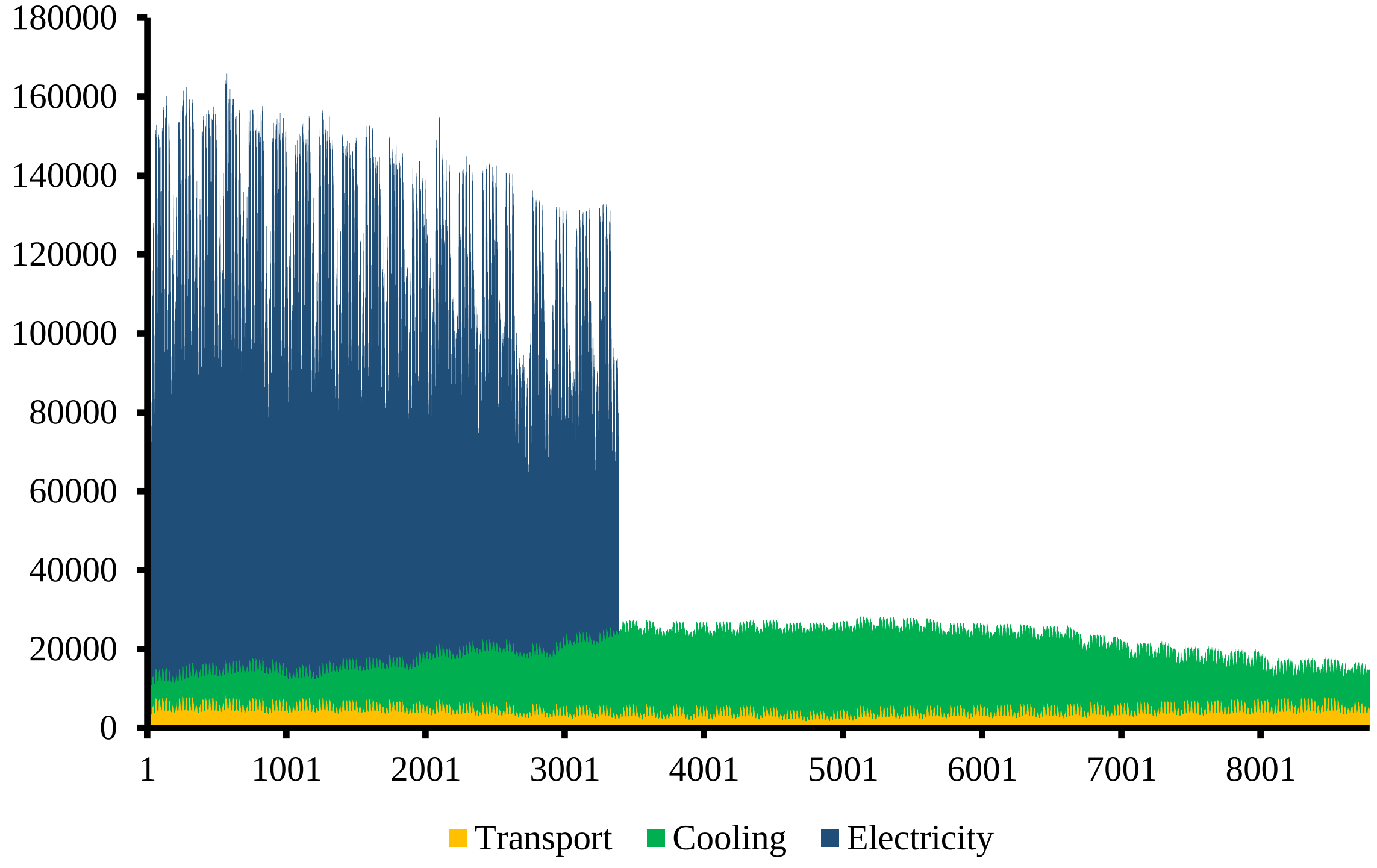

Fig. 2: 2050 total electricity demand for Nigeria

## 2.5. Machine learning modeling

### 2.5.1. Overview of machine learning modeling

ML offers an approach for models to learn specific relationships from data without being programmed to do so. In this study, the dataset used to train the ML models comprised of inputs generated using CCD, and the outputs obtained from the EnergyPLAN in-silico experimentation. Three machine learning models were employed in an attempt to identify the best models that described the two scenarios: WoEST and WEST, including XGBoost, SVR, and MLP. All data-preprocessing and modeling steps were implemented using the Python programming language (version 3.11.5). The ML models were implemented using the Sci-kit learn library. Due to the varying scales of the input features, they were scaled using z-score normalization, implemented using the Sci-Kit Learn library, and described in Equation (2), where $z$ is the normalized value, $x$ is the original value of the feature, $\mu$ is the mean of the feature, and $\sigma$ is the standard deviation of the feature. This centers the data around 0 and scales it to have a standard deviation of 1. This technique gives all features equal chances to have an effect in predicting the model output.

$$z = \frac{x - \mu}{\sigma} \quad (2)$$

To ensure robust model evaluation, a k-fold cross-validation method was utilized, which is crucial for minimizing overfitting and enhancing model generalization. This technique divides the dataset into k subsets, trains the model on k-1 subsets, and validates it on the remaining one. This process is repeated k times, each with a different validation set. Specifically, a 5-fold cross-validation was employed, with a 70:30 train-test split, meaning 70% of data was used for training and 30% for testing. Cross-validation was performed on the entire training set, and testing was done on an unseen test set.

During training, the ML models' hyperparameters were optimized using Bayesian optimization, implemented using the Optuna library developed by Akiba et al. (2019). Bayesian optimization facilitates continuous model improvement across iterations by compounding on insights from previous iterations to intelligently tune hyperparameter values, thereby enhancing the likelihood of achieving desired outcomes in subsequent iterations. This methodology stands superior to conventional approaches like grid search and random search, which lack the adaptive learning capability exhibited in Bayesian optimization (Li & Kanoulas, 2018).

### 2.5.2. Extreme gradient boosting

XGBoost is a highly efficient and scalable implementation of gradient boosting, widely used for its superior performance in modeling tasks (Chen & Guestrin, 2016). At its core, XGBoost builds an ensemble of decision trees sequentially, where each tree is trained to correct the errors of its predecessors. The prediction at step $t$ for a given input is given in Equation (3).

$$\hat{y}_i^{(t)} = \emptyset + \sum_{k=1}^{t} f_k(x_i) \quad (3)$$

Where, $\emptyset$ is the initial prediction and $f_k$ is the $k\text{-}th$ tree.

Equation (4) shows the objective function minimized by XGBoost, which consists of a loss function $L$ and a regularization term $\Omega$:

$$L^{(t)} = \sum_{i=1}^{n} L\left(y, \hat{y}^{(t)}\right) + \sum_{k=1}^{t} \Omega(f_k) \quad (4)$$

The loss function $L$ measures the difference between the predicted values $\hat{y}^i$ and the actual target values $y^i$, and the regularization term $\Omega(f_k)$ controls the complexity of the model to prevent overfitting (Fan et al., 2018). The model iteratively reduces residuals, which are the differences between actual and predicted values. At each step, a new tree is added to model these residuals as depicted in Equation (5).

$$\hat{y}_i^{(t)} = \hat{y}_i^{(t-1)} + f_t(x_i) \tag{5}$$

### 2.5.3. Support vector regression

SVR is a powerful and versatile regression technique based on the principles of Support Vector Machines (SVMs). The SVM model performs regression by using a set of kernel functions that implicitly transform the original, lower-dimensional input data into a higher-dimensional feature space (Fan et al., 2018). SVM aims to find a function that approximates the mapping between input features and target values while maintaining a balance between model complexity and prediction accuracy (Smola & Schölkopf, 2004). The goal of SVR is to find a function $f(x)$ that deviates from the actual target values by a value no greater than $\varepsilon$ for each training point, while also being as flat as possible. This is formulated as shown in Equation (6):

$$f(x) = \omega\varphi(x) + b \tag{6}$$

where $\omega$ is the weight vector, b is the threshold, and $\varphi(x)$ represents the higher-dimensional feature space transformed from the input vector $x$.

The optimization problem for SVR is as outlined in Equations (7) and (8).

$$minimize \quad \frac{1}{2}\|\omega\|^2 + C\sum_{i=1}^{l}(\xi_i + \xi_i^*) \tag{7}$$

$$subject\ to \begin{cases} y_i - \langle\omega, x_i\rangle - b \le \varepsilon + \xi_i \\ \langle\omega, x_i\rangle + b - y_i \le \varepsilon + \xi_i^* \\ \xi_i, \xi_i^* \qquad\qquad \ge 0 \end{cases} \tag{8}$$

Here, $\xi_i$ and $\xi_i^*$ are slack variables that allow for deviations beyond $\varepsilon$ (Drucker et al., 1996), and $C$ is a regularization parameter that determines the trade-off between the flatness of $f$ and the amount up to which deviations larger than $\varepsilon$ are tolerated (Smola & Schölkopf, 2004).

The kernel trick allows SVR to perform in high-dimensional feature spaces efficiently. By using a kernel function, the dot product $\langle x_i, x_j\rangle$ can be replaced with $K\langle x_i, x_j\rangle$, enabling the algorithm to capture complex, non-linear relationships (Scholkopf et al., 1999).

### 2.5.4. Multi-layer perceptron

MLP is a type of artificial neural network (ANN) architecture that consists of multiple layers of neurons, typically organized into an input layer, one or more hidden layers, and an output layer (Zhou et al., 2021). MLP models are widely used for their ability to model complex, non-linear relationships in data. Each neuron in an MLP performs a weighted sum of its inputs and passes the result through a non-linear activation function. Mathematically, for a neuron $j$ in layer $l$, the output $h_j^{(l)}$ is given by Equation (9).

$$h_j^{(l)} = \sigma\left(\sum_{i=1}^{n^{(l-1)}} \omega_{ij}^{(l)} h_i^{(i-1)} + b_j^{(l)}\right) \tag{9}$$

Where, $\omega_{ij}^{(l)}$ are the weights connecting neuron $i$ in layer $l-1$ to neuron $j$ in layer $l$, $b_j^{(l)}$ is the bias term for neuron $j$ in layer $l$, and $\sigma$ is the activation function. The output of the MLP for a given input $\boldsymbol{x}$ is obtained by propagating the input through the network layers, with the final output layer producing the predicted values.

MLPs are trained using a supervised learning approach, where the objective is to minimize the difference between the predicted values and the actual target values. This is achieved by minimizing a loss function $L$, such as the Mean Squared Error (MSE) for regression tasks depicted in Equation (10), where $y$ is the actual target value, $\hat{y}$ is the predicted target value, and $n$ is the number of examples.

$$L(y, \hat{y}) = \frac{1}{n}\sum_{i=1}^{n}(y_i - \hat{y}_i)^2 \tag{10}$$

This is typically done using the backpropagation algorithm, which consists of two phases, a forward pass and backward pass.

During the forward pass, the output of the network for a given input is computed by propagating the input through the layers. In the backward pass, the gradient of the loss function with respect to each weight and bias is computed using the chain rule, and the weights and biases are updated using gradient descent or its variants. The weight update rule for a weight $\omega_{ij}^{(l)}$ at iteration $t$ is as shown in Equation (11).

$$\omega_{ij}^{(l)}(t+1) = \omega_{ij}^{(l)}(t) - \eta \frac{\partial L}{\partial \omega_{ij}^{(l)}} \tag{11}$$

Where, $\eta$ is the learning rate, a hyperparameter that controls the step size of the updates.

## 2.6. Optimization of electricity scenarios

Metaheuristic algorithms are a class of optimization techniques that are designed to find good solutions for complex problems where traditional methods may be impractical. These algorithms are typically inspired by natural processes and phenomena. They are characterized by their ability to explore a large search space and escape local optima, making them suitable for solving problems with many local optima. To derive the optimal set of input capacities that minimized total annual cost, $CO_2$ emission, CEEP, and electricity import, a multi-objective optimization was carried out, employing genetic algorithm (GA). GA is an evolutionary-based type of metaheuristic algorithm with a key characteristic being its independence from derivatives within the search space, unlike traditional mathematical optimization methods (Zhao et al., 2020). It is a prominent type of metaheuristic algorithm inspired by the principles of natural selection and genetics. GA operates on a population of candidate solutions, called individuals, each represented by a chromosome. The basic idea is to evolve the population over several generations to improve the quality of the solutions (Amenaghawon et al., 2024). In the context of optimizing an objective function, each chromosome represents a set of parameters for the objective function. The fitness function is typically based on the value of the objective function, where the goal is to maximize or minimize this value, depending on the problem at hand.

Validation of the ML models was done by using the optimized capacities gotten through GA optimization as inputs to the EnergyPLAN software, and comparing the GA minimized outputs with the EnergyPLAN outputs.

## 2.7. Feature ranking

To understand the decision-making process of ML models and to gain more insight into the modeled system, the SHapley Additive exPlanations (SHAP), a unified approach to interpreting model predictions, was employed using the SHAP library in Python programing language. SHAP leverages game theory to explain the output of any machine learning model by assigning each feature an importance value for a particular prediction (Lundberg et al., 2017). The SHAP value $\emptyset$ for a feature $i$ is calculated using the Shapley value from cooperative game theory as delineated in Equation (12), where $F$ is the set of all features, $S$ is a subset of features, and $f_S(x_S)$ is the model prediction considering only features in $S$.

$$\emptyset_i = \sum_{S \subseteq F \setminus \{i\}} \frac{|S|!\,(|F| - |S| - 1)!}{|F|!} \left[ f_{S\cup\{i\}}\left(x_{S\cup\{i\}}\right) - f_S(x_S) \right] \tag{12}$$

SHAP waterfall plots provide a visual breakdown of the contribution of each feature to the model's prediction for a single instance. This plot starts from the baseline value, which is the average model output over the training dataset. Each feature's contribution is then added (or subtracted) step-by-step, showing how the model's prediction deviates from the baseline. The waterfall plot thus helps in understanding how individual feature values lead to a final single-value prediction. In the waterfall plot, the SHAP value $\emptyset_i$ for each feature $i$ is depicted, showing how the feature shifts the model output either towards or away from the baseline. The SHAP beeswarm plots provide a summary of the SHAP values for all features across the entire dataset, giving an overview of feature importance and their impact distribution. Each point on the bee swarm plot represents a SHAP value for a feature and an instance. The color of the points indicates the feature value (often from low to high), and the spread of points along the x-axis shows the distribution of the SHAP values. Beeswarm plots help identify which features are most influential overall and how they interact with the model's predictions. Features with large spreads are more influential, and the direction of their SHAP values indicates whether higher values of the feature lead to higher or lower model predictions.

## 3. Results and Discussion

### 3.1. Tuning of hyperparameters

The models' hyperparameters were tuned via a combination of a 5-fold cross-validation technique and Bayesian optimization. All three ML models were used to predict the four responses and the hyperparameter configurations are shown in Table S1-S6 in supplementary materials. However, **Error! Reference source not found.** shows the optimized hyperparameters for the models which had the best performances for both scenarios. The optimized hyperparameters were then used in conjunction with their corresponding models to predict the responses.

Table 5: Optimized hyperparameters for selected models in both scenarios: WoEST and WEST

| | WoEST Scenario | | | WEST Scenario | | |
|---|---|---|---|---|---|---|
| Output | Model | Hyperparameter | Value | Model | Hyperparameter | Value |
| | SVR | C | 889908.6056 | SVR | C | 938266.6722 |
| Total annual cost | | Epsilon | 0.11778 | | Epsilon | 7.6347 |
| | | Kernel | rbf | | Kernel | rbf |
| | | Degree | 1 | | Degree | 4 |
| | | Gamma | 0.0175 | | Gamma | 0.0088 |
| $CO_2$ emission | SVR | C | 567754.8073 | SVR | C | 71459.2769 |
| | | Epsilon | 0.0492 | | Epsilon | 0.9032 |
| | | Kernel | rbf | | Kernel | rbf |
| | | Degree | 4 | | Degree | 1 |
| | | Gamma | 0.0029 | | Gamma | 0.0036 |
| CEEP | SVR | C | 653379.5949 | XGB | Max depth | 5 |
| | | Epsilon | 0.0068 | | Learning rate | 0.2119 |
| | | Kernel | rbf | | Number of estimators | 531 |
| | | Degree | 1 | | Minimum child weight | 1 |
| | | Gamma | 0.0081 | | Gamma | 0.8354 |
| | | | | | Subsample | 0.8560 |
| | | | | | Column Sample by Tree | 0.9500 |
| | | | | | Alpha | 0.0614 |
| | | | | | Lambda | 0.9728 |
| | | | | | Random state | 599 |
| Import | MLP | Number of hidden layers | 4 | MLP | Number of hidden layers | 4 |
| | | Hidden layer 1 units | 33 | | Hidden layer 1 units | 7 |
| | | Hidden layer 2 units | 10 | | Hidden layer 2 units | 28 |
| | | Hidden layer 3 units | 8 | | Hidden layer 3 units | 23 |
| | | Hidden layer 4 units | 40 | | Hidden layer 4 units | 19 |
| | | Activation | relu | | Activation | relu |
| | | Alpha | 0.2007 | | Alpha | 0.9366 |
| | | Solver | lbfgs | | Solver | lbfgs |
| | | Random state | 52 | | Random state | 72 |

### 3.2. Performance of ML models

The accuracies of the optimized ML models were assessed using the R-squared ($R^2$) value on the test data set, as shown in Table 6. The $R^2$ value quantifies the proportion of variance in the dependent variable that is predictable from the independent variables. Remarkably, all models exhibited high performances across all four responses, underscoring the reliability of machine learning algorithms in modeling energy systems (Forootan et al., 2022). This could be attributed to the ability of these models to map nonlinear relationships between the data and labels.

Table 6: Summary of the R-squared values for each model across the four output features

| | WoEST Scenario | | | | WEST Scenario | | | |
|---|---|---|---|---|---|---|---|---|
| Model | Total annual cost | $CO_2$ emission | CEEP | Electricity import | Total annual cost | $CO_2$ emission | CEEP | Electricity import |
| SVR | 0.99967 | 0.99940 | 0.99630 | 0.99886 | 0.99906 | 0.99877 | 0.92728 | 0.99929 |
| XGB | 0.99863 | 0.99598 | 0.99305 | 0.99386 | 0.99585 | 0.99875 | 0.99220 | 0.99986 |
| MLP | 0.99875 | 0.99680 | 0.98972 | 0.99981 | 0.99349 | 0.99807 | 0.98361 | 0.99988 |

### 3.3. Energy systems optimization

GA was applied for a multi-objective optimization of the ML models for Nigeria's 2050 electricity scenarios, with and without grid-scale electricity storage. The goal of this multi-objective optimization was to minimize total annual cost, $CO_2$ emission, CEEP, and electricity import. As shown in Fig. 3 and Fig. 4 which show the optimization plots for the scenarios without and with storage technologies, respectively, GA was generally efficient in minimizing the outputs with increasing iterations. Table 7 shows the configurations of the GA for both electricity scenarios.

Table 7: Optimization configuration for genetic algorithm

| | WoEST Scenario | WEST Scenario |
|---|---|---|
| Parameter | Value | Value |
| Population size | 20.000 | 20.000 |
| Epoch | 200.000 | 200.000 |
| Crossover probability | 0.500 | 0.500 |
| Mutation Probability | 0.001 | 0.001 |

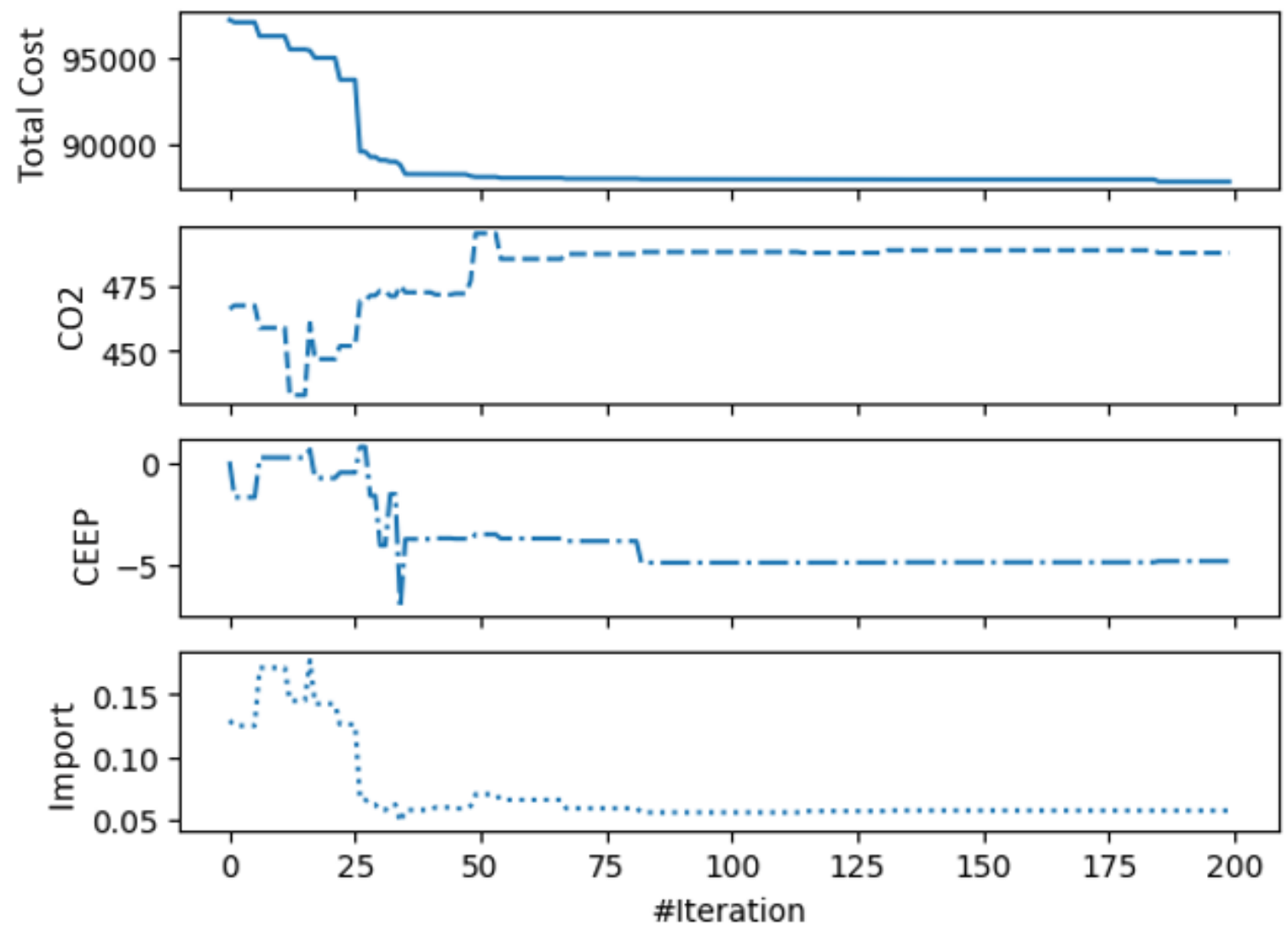

Fig. 3: Multi-objective optimization plot for WoEST scenario

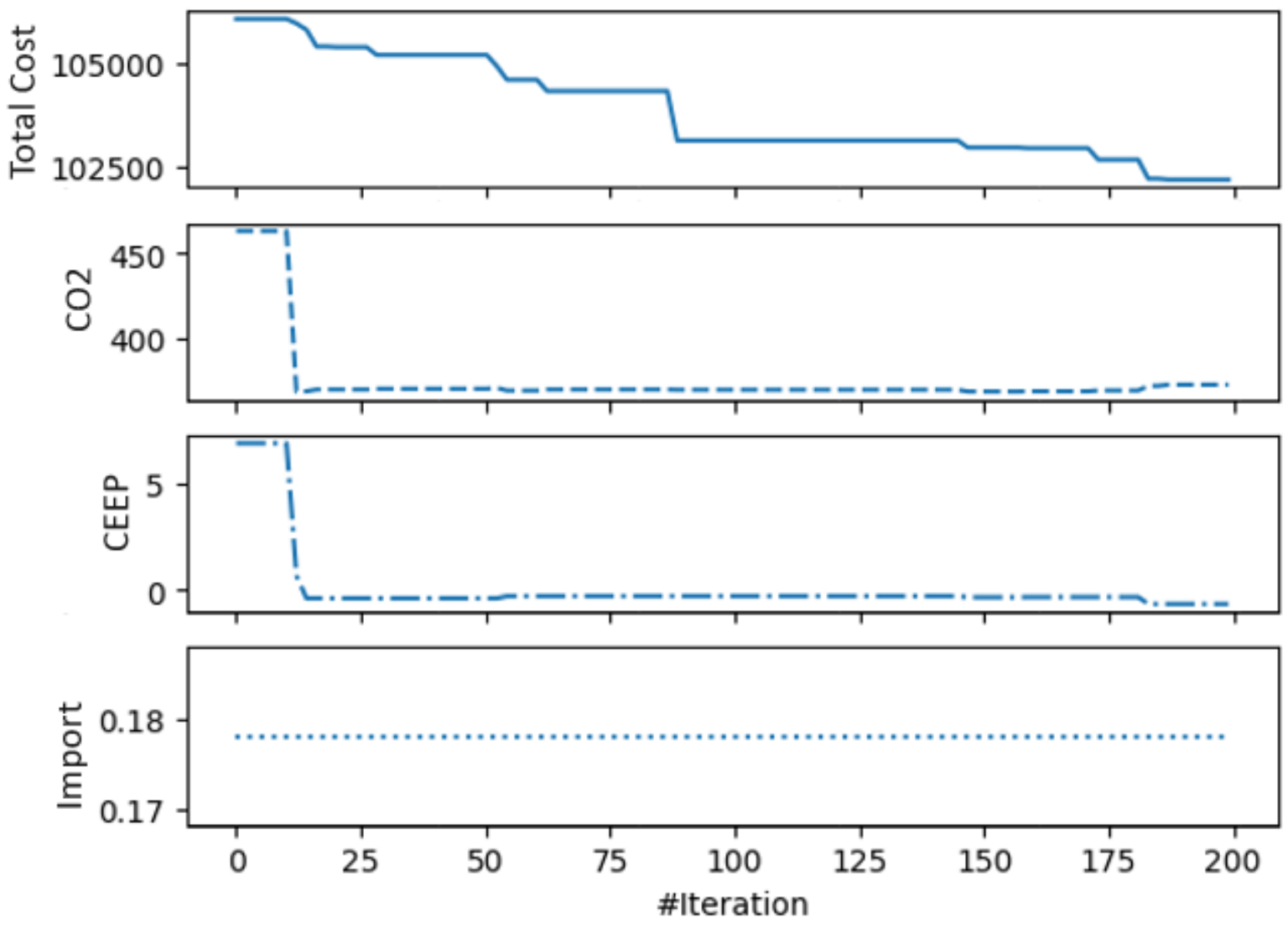

Fig. 4: Multi-objective optimization plot for WEST scenario

The optimized capacities which gave the minimum outputs are shown in Table 8 for both scenarios. These capacities were further used as inputs in the EnergyPLAN software for ML model validation. The optimized GA outputs as well as the EnergyPLAN validated results are presented in Table 9 and Table 10, for the WEST and WoEST scenarios, respectively.

Table 8: Optimized input capacities for WoEST and WEST scenarios

| Variable | Value | |
|---|---|---|
| | WoEST Scenario | WEST Scenario |
| PP | 149038.063 | 124451.68 |
| Nuclear | 14800.0681 | 27206.547 |
| Wind | 7800.56551 | 9599.6358 |
| PV | 33726.5482 | 78492.168 |
| RH | 11200.0441 | 17850.746 |
| CSP | 846.001487 | 305.43339 |
| G2B | N/A | 26099.6708 |
| VBS | N/A | 33.003883 |
| B2G | N/A | 27420.409 |
| PHS CC | N/A | 49888.164 |
| PHS DC | N/A | 44477.028 |
| PHS SC | N/A | 19.056796 |
| BESS CC | N/A | 60657.83 |
| BESS DC | N/A | 67707.482 |
| BESS SC | N/A | 1240.74276 |

Table 9: ML and EnergyPLAN-validated outputs for WEST scenario

| | Total annual cost | $CO_2$ emission | CEEP | Electricity import |
|---|---|---|---|---|
| ML | 102192 | 373 | 0 | 0 |
| EnergyPLAN | 106750 | 363 | 0 | 0 |

Table 10: ML and EnergyPLAN-validated outputs for WoEST scenario

| | Total annual cost | $CO_2$ emission | CEEP | Electricity import |
|---|---|---|---|---|
| ML | 87850 | 488 | -5 | 0 |
| EnergyPLAN | 102972 | 449 | 0 | 0 |

As shown in Table 9 and Table 10, the ML algorithms were effective at modeling the EnergyPLAN simulations. The similarity between the results obtained from the EnergyPLAN model and those obtained from the multi-objective optimization validates the ML modelling and optimization carried out. The slight disparities between the two results can, however, be attributed to the fact that rather than establishing a series of balance equations that are solved numerically as in optimization models, EnergyPLAN is based on a series of endogenous priorities within, and pre-defined procedures for simulating the operation of units (Lund et al., 2021). The optimization model employs a stochastic method, while EnergyPLAN's approach is entirely deterministic and devoid of any stochastic components.

### 3.4. Comparison of the two scenarios, WoSET and WEST

As shown in Fig. 5, the total annual cost had a 3.67% increase from 102,972 MUSD in the WoEST scenario, to 106,750 MUSD in the WEST scenario. This increase can be attributed to the additional costs associated with implementing storage technologies within the energy system (Kittner et al., 2020; Ozdemir et al., 2016; Zakeri & Syri, 2015). This raises questions about the economic viability of grid-level storage systems. However, the overall benefits of storage technologies in the electricity grid may exceed its costs. $CO_2$ emission reduced by 19.14% from 449 Mt in the WoEST scenario to 363 Mt in the WEST scenario. This cut in $CO_2$ emission helps Nigeria in achieving the 2050 climate vision, to reduce the current emission by 50% (Department of Climate Change, 2021). This reduction can be primarily attributed to the improved RES penetration enabled by energy storage technologies, which facilitate the displacement of fossil fuel-based generation and inherently leads to lower carbon emissions. This trend is generally observed. Wang et al., (2018) found that higher capacities of energy storage technologies increase the utilization efficiency of renewable energy, subsequently reducing carbon emissions in Chongming, China. Also, according to Krajačić et al., (2011), the use of renewable energy sources combined with energy storage could reduce $CO_2$ emissions in Croatia by 82% by 2050. CEEP and electricity import remained zero in both scenarios as the scenarios presented represent the best-case scenarios given the capacities presented, where there is no excess electricity produced and no reliance on other countries for electricity. This indicates that the ML and multi-objective optimization algorithms effectively minimized both CEEP and electricity imports, optimizing the system to avoid excess electricity and dependency on external sources. This suggests that both scenarios provided sufficient electricity to meet demand, with variable sources balanced precisely to avoid overproduction.

The use of storage technologies allowed for a 37% increase in the integration of RES, without leading to CEEP; from 131.24 TWh of RES electricity production without electricity storage technologies, to 179.74 TWh of RES electricity production with electricity storage technologies incorporated in the electricity grid. Strategic planning is therefore essential to lower the overall cost of integrating energy storage into the grid. Pumped hydro storage could be utilized more due to its generally lower cost and long lifespan of approximately 50 years, compared to batteries, which have high costs and shorter life spans, as shown in Table 3. Battery costs are, however, expected to decrease as researchers continue to investigate novel battery technologies for lower cost, and increased energy density (Faunce et al., 2018; Kittner et al., 2020; Mongird et al., 2020; Tianquan et al., 2015).

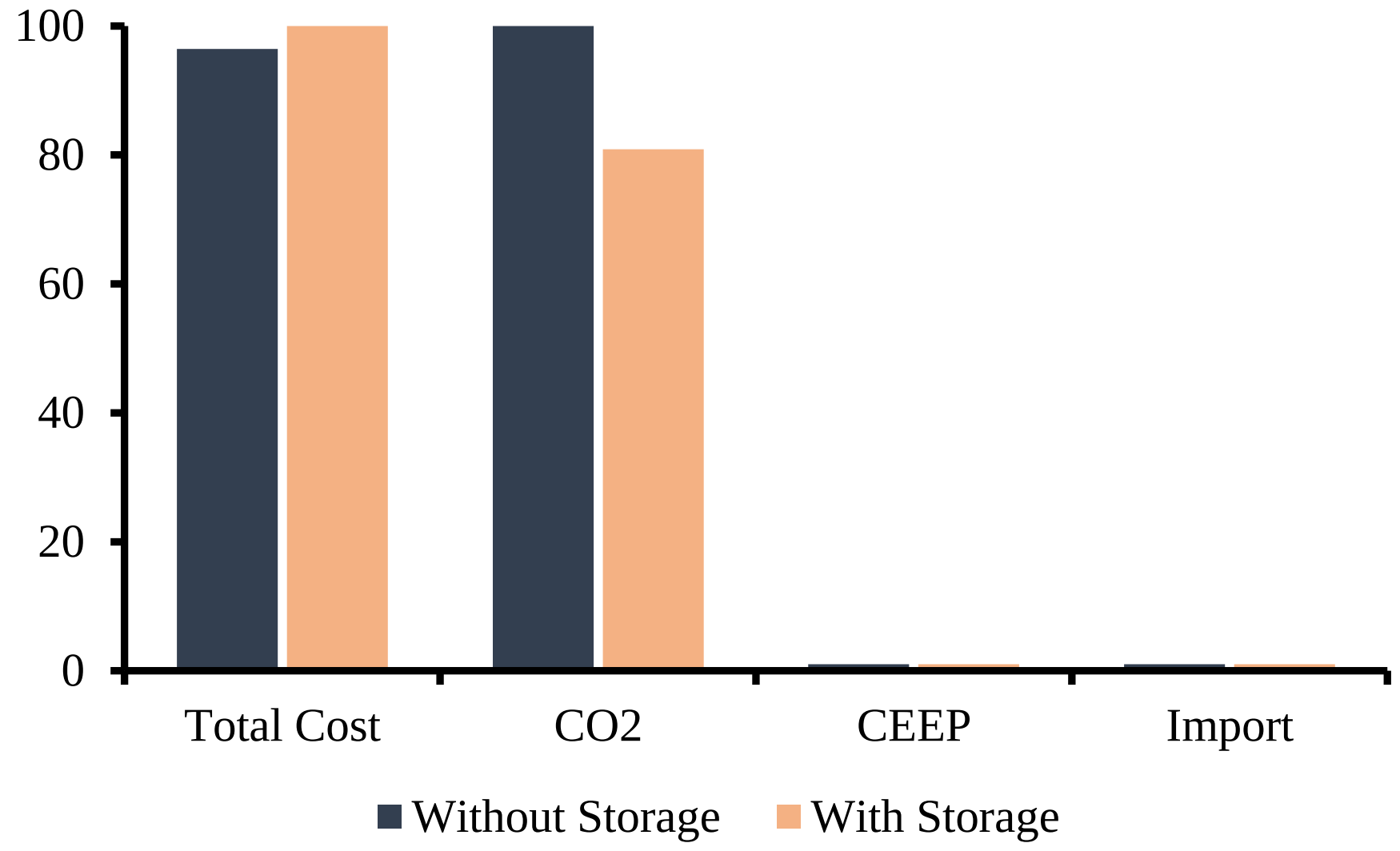

Fig. 5: Percentage change of outputs in both scenarios with and without storage

Given the two best-case scenarios, with and without storage technologies, it is seen that each has its merit and demerit. While a scenario with electricity storage is 3.67% more costly than a system without energy storage, it however, produces 19.14% less $CO_2$ emission than the scenario without storage. Given the global drive towards the reduction of carbon emission, storage technologies present a good option for the electricity system. Furthermore, the use of storage technologies allows for the use of 37% more RES, expanding a diversification of the electricity mix and reducing the dependence on fossil fuel. Given the ongoing research in the field of electricity storage technologies, the high cost associated with storage technologies is expected to reduce in the coming years, making it an even better scenario that fits into both the environmental and economic needs of Nigeria's electricity landscape.

### 3.5. Model interpretation and analysis for energy storage scenario

Incorporating energy storage into the energy system is desirable because it allows for the wide incorporation of VRES (Zerrahn & Schill, 2017), reduces dependence on fossil fuel, and lowers carbon emission. Understanding the effects of each independent variable on the system enhances the understanding of the system and enables resource prioritization to meet a desired objective.

Furthermore, this gives insight into the decision-making process of the machine learning model and elucidates the importance of the various input features to enhance the understanding of machine learning models.

In this subsection, the effects of the different input features on the four target variables, total annual cost, $CO_2$ emission, CEEP and electricity import were elucidated using waterfall and beeswarm plots. These visualizations were created using the SHAP library in the Python programming language. Contour plots were also used to show the interaction effects between the most significant inputs on the outputs, implemented using the matplotlib library.

As shown in Fig. 6, different independent variables affect the different models in varying degrees. The extents of these effects were described using the Shapley values, computed and visualized using the SHAP library. It can be seen from Fig. 6 that PP is a potent variable, affecting total annual cost, $CO_2$ emission, and import to the largest extents. As shown in Fig. 6 and Fig. 7, while higher PP capacities tend to favor lower cost, and lower electricity import, it has the most negative effect on $CO_2$ emission. This is because the deployment of PP is typically associated with the use of fossil fuel, which are among the largest contributors of $CO_2$ emission (Paraschiv & Paraschiv, 2020). It can also be seen from Fig. 6(a) that higher CSP and BESS SC tend to lead to higher costs. This could be due to installation and maintenance costs of these technologies as shown in Table 3. As shown in Fig. 6(b), $CO_2$ emission is driven solely by PP, while all other technologies which include nuclear, RES, and storage technologies act to reduce $CO_2$ emission. Fig. 6(c) shows that RES are the major factors which drive CEEP, this excess electricity can, however, be reduced with the use of storage technologies, as these are shown to reduce CEEP from RES (Xiao et al., 2024). High PV capacities result in an increase in CEEP, whereas higher BESS, PHS, and EV capacities lead to reduced CEEP (Ozdemir et al., 2016; Vaziri Rad et al., 2023). This is because, high PV capacities lead to significant amount of electricity generation when solar irradiance is high, which can exceed the immediate demand and grid capacity, leading to higher CEEP. Conversely, energy storage systems mitigate this issue by storing (charging) excess electricity during peak generation periods and releasing (discharging) it back to the grid during periods of high demand or low generation, thereby reducing CEEP.

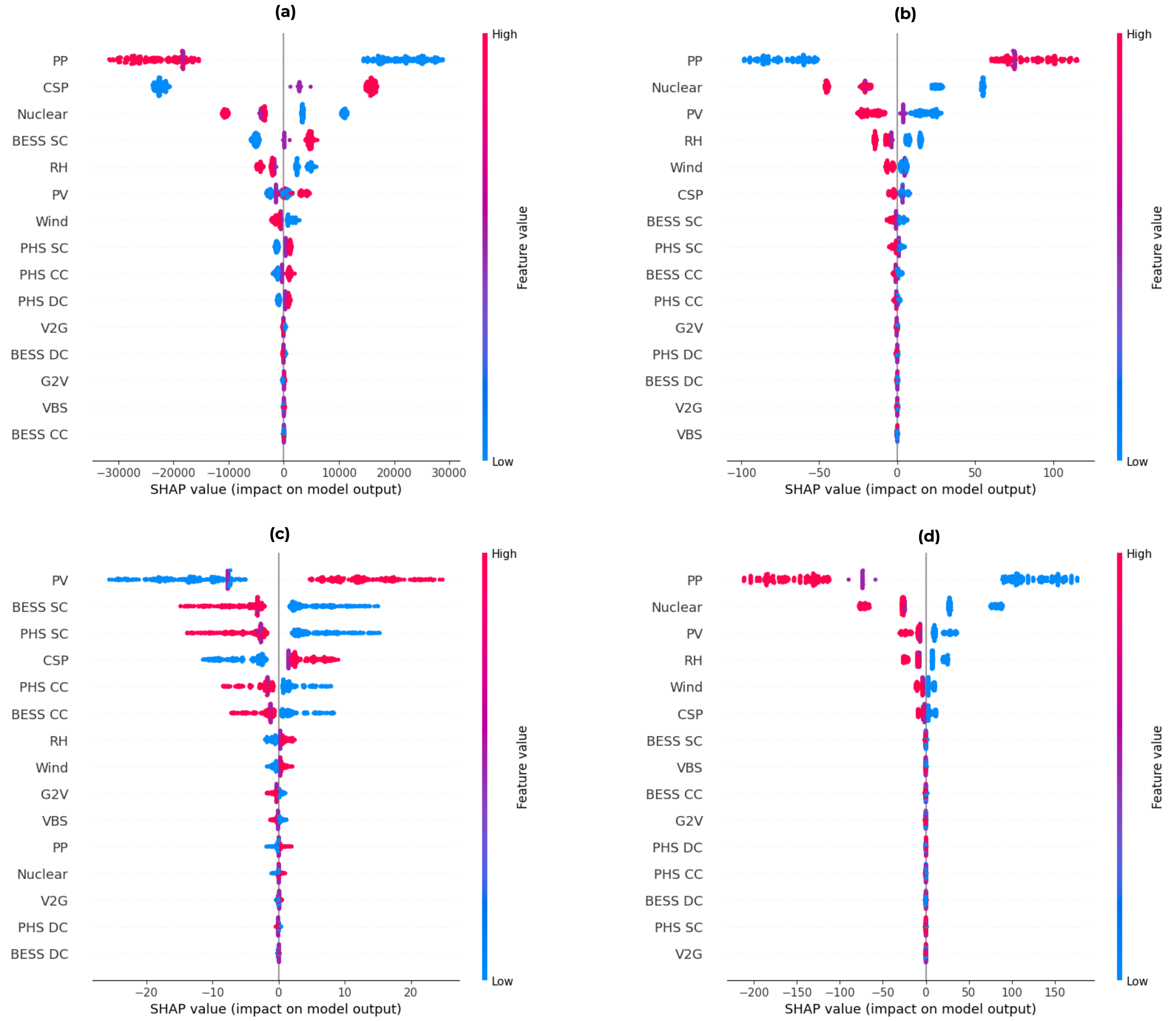

Fig. 6: Beeswarm plots showing features importance for (a) total annual cost (b) $CO_2$ emission (c) CEEP (d) electricity import

This underscores the need for the development of electricity storage systems. In effect, RES are needed for reduced reliance on $CO_2$, and storage technologies are needed to optimally utilize RES (Bellocchi et al., 2019; Ozdemir et al., 2016). Furthermore, it is seen from Fig. 6(d) that all power generating sources favor lower electricity import. This justifies the need for a diversification of the electricity landscape to reduce reliance on other countries for electricity. It is also seen that low PP capacities has the most effect on increasing the reliance on electricity imports. This is because the installed RES capacities may not generate sufficient power to meet demand. Hence, fossil power plants are required to compensate for this shortfall.

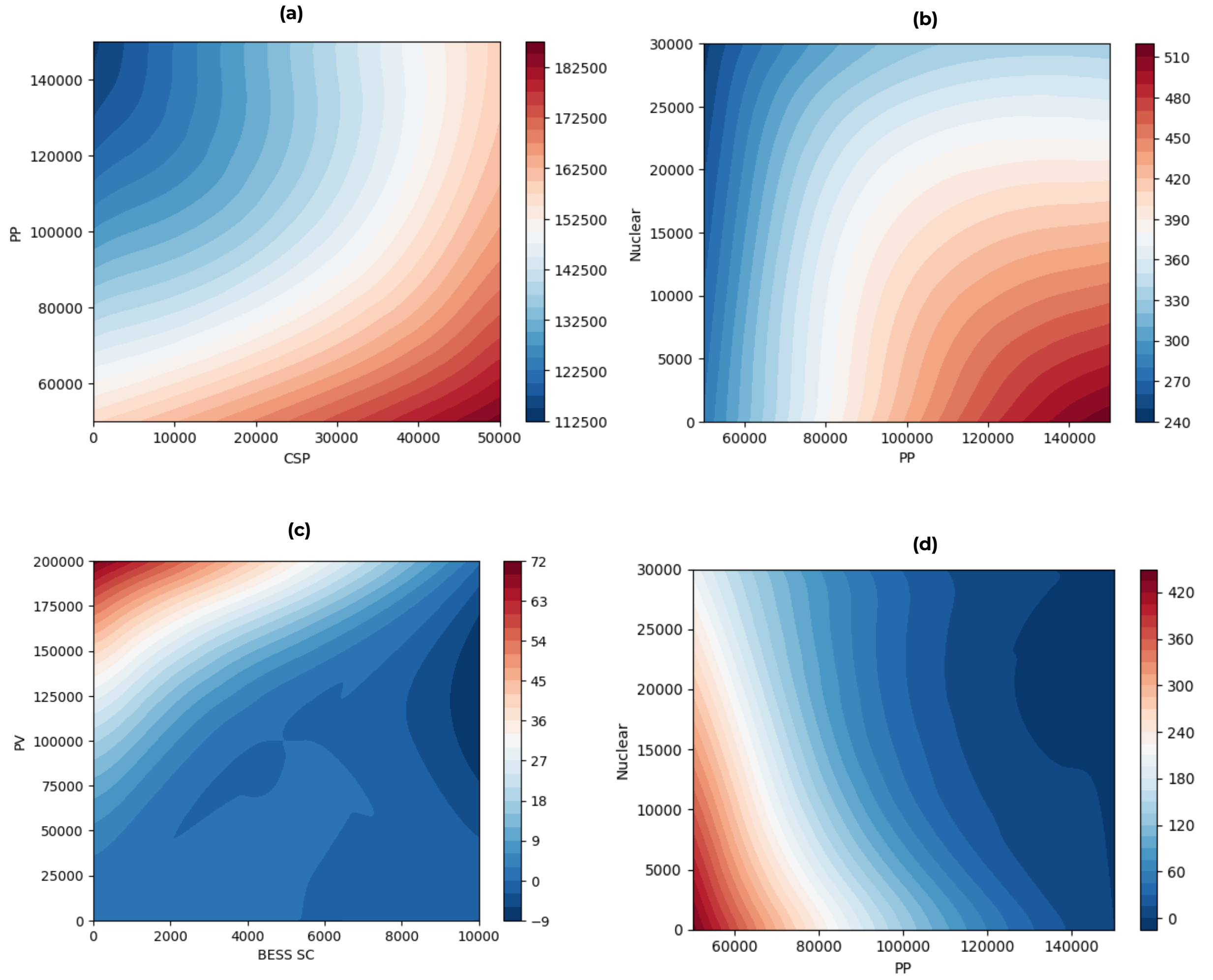

Fig. 7: Contour plots for (a) total annual cost (b) $CO_2$ emission (c) CEEP (d) electricity import

Although it is desirable to eliminate the usage of fossil fuel-based power production, solely relying on renewable energy technologies to satisfy energy demand without resorting to electricity imports may not be feasible in the near future. In a study by Bamisile et al. (2020), they considered 11 combinations of RES to meet Nigeria's electricity demand by 2030 and concluded that none of these combinations were sufficient without fossil fuel power plants.

The effects of the optimal independent variables presented in **Error! Reference source not found.** on the minimized outputs were further analyzed using Shapley values and visualized through waterfall plots, as shown in Fig. 8. This comprehensive approach provides a step-by-step visualization of how each variable contributes to moving the prediction from the baseline (mean) value to the final prediction. This allows the evaluation of not only the magnitude but also the direction of the influence of each variable. By using Shapley values, the contribution of each independent variable to the minimized output can be quantified, thereby elucidating the relative importance and impact of each variable on the optimization goals.

Fig. 8(a) shows that the optimized total annual cost of 102,192 MUSD as shown in Table 9 was significantly impacted by an optimal CSP capacity of 305.433 MW, which minimized total annual cost from a higher mean. This implies that the optimal CSP capacity is within the low region shown in Fig. 6(a). Fig. 8(a) further shows that all the independent variables had a positive effect on minimizing total annual cost. Moreover, Fig. 8(b) highlights the optimized values for $CO_2$ emissions, showing that a nuclear capacity of 27,206.547 MW was crucial in achieving a minimized $CO_2$ output of 373 Mt. Conversely, the analysis reveals that a PP capacity of 120,451.678 MW is associated with an increase in $CO_2$ emissions. This correlation suggests that higher PP capacities, often relying on natural gas or coal-fired power plants for quick ramp-up, contribute to higher emission levels. This effect was, however, cancelled out by the optimized nuclear capacity which had a higher absolute Shapley value of 35.03. Nuclear energy, known for its low carbon footprint compared to fossil fuels, evidently contributes significantly to emission reduction targets. This finding underscores the importance of nuclear power in a diversified energy mix aimed at sustainable and low-emission energy production. This insight emphasizes the need for balancing peak demand with cleaner energy sources or enhancing energy storage solutions to mitigate the environmental impact.

The Shapley value analysis and waterfall plots not only provide a transparent and interpretable method for understanding the effects of independent variables on the optimized outcomes but also highlight the critical role of specific energy capacities in cost and emission minimization. These insights can guide policymakers and stakeholders in making informed decisions about energy planning and investment, fostering a more sustainable and economically viable power system.

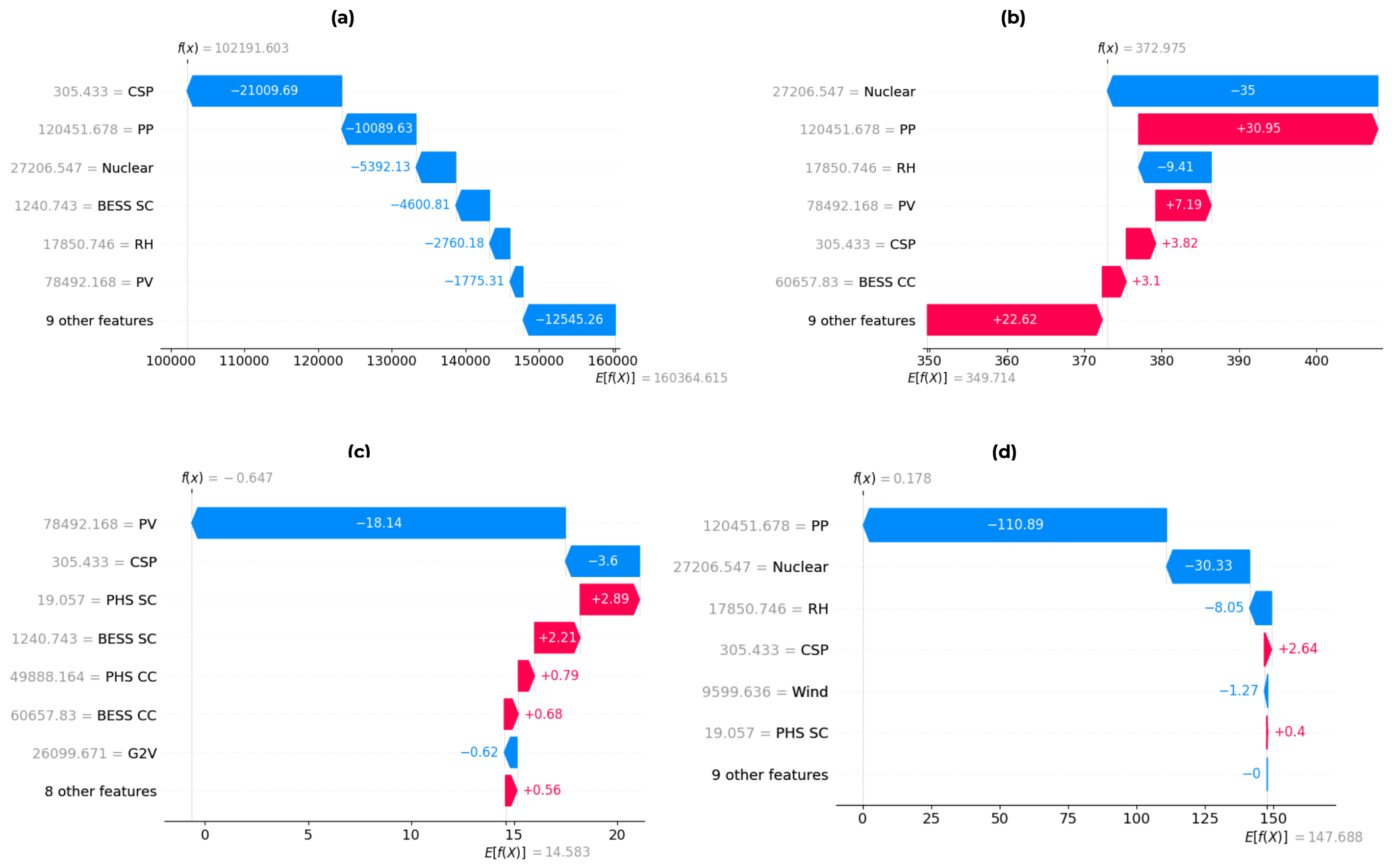

Fig. 8: Waterfall plots of optimized system for (a) total annual cost (b) $CO_2$ emission (c) CEEP (d) electricity import

## 4. Conclusion

This research presents a comprehensive assessment of Nigeria's future electricity scenarios by 2050, focusing on the impacts of incorporating grid-scale electricity storage technologies. Using a systematic approach that includes modeling and data analysis, two primary scenarios were explored: with and without electricity storage technologies. Our findings indicate that while the inclusion of energy storage technologies in Nigeria’s energy landscape by 2050 leads to a higher overall cost, it also facilitates the integration of more renewable electricity sources, significantly reducing $CO_2$ emissions. This underscores the trade-off between economic expenditure and environmental impact. However, given the global drive towards reducing carbon emissions and ongoing research to reduce the cost of storage and RES technologies, an electricity system with storage technologies presents a viable option for Nigeria’s future electricity landscape.

Specifically, machine learning algorithms offer a reliable approach for modeling Nigeria’s energy system, enabling the prediction of future variables. A multi-objective optimization approach proved to be efficient, as it ensures a systematic exploration of different variables affecting the energy system, leading to robust and generalizable findings. Fossil power plants were effective in reducing both total annual costs and electricity imports, but they also contributed significantly to $CO_2$ emissions, highlighting the need for a balanced approach in their deployment. RES and nuclear power were effective in reducing $CO_2$ emissions. However, RES also led to an increase in CEEP, necessitating careful management to ensure grid stability and efficiency. The use of electricity storage technologies proved crucial in mitigating CEEP, helping balance supply and demand and enhancing the integration of renewable energy sources into the grid. A BESS capacity of 1240.74 GWh, a PHS capacity of 19.05 GWh, and an EV storage capacity of 33 GWh are needed for zero CEEP, allowing for a 37% increase in RES share, a 19.14% decrease in $CO_2$ emissions, with a 3.67% increase in total annual cost.

This research demonstrates that the strategic inclusion of electricity storage technologies in Nigeria's energy mix for 2050 can significantly reduce $CO_2$ emissions and manage excess electricity production, albeit at a higher cost. The findings emphasize the importance of a multi-faceted approach that includes a mix of fossil, renewable, and nuclear energy sources, complemented by advanced storage solutions, to achieve a sustainable and efficient electricity system. Future work should focus on using distribution, demand, and cost data from various sources to create a more holistic picture of how electricity storage technologies will affect Nigeria's energy landscape.

**CRediT authorship contribution statement**

**Stanley Aimhanesi Eshiemogie:** Writing – review & editing, Writing – original draft, Visualization, Validation, Software, Project administration, Methodology, Investigation, Formal analysis, Data curation, Conceptualization. **Peace Precious Aielumoh:** Writing – review & editing, Writing – original draft, Software, Investigation, Formal analysis, Data curation, Conceptualization. **Tobechukwu Okamkpa:** Writing – review & editing, Writing – original draft, Software, Investigation, Formal analysis, Data curation, Conceptualization. **Miracle Chinonso Jude:** Writing – original draft, Conceptualization. **Lois Efe:** Writing – review & editing. **Andrew Nosakhare Amenaghawon:** Writing – review & editing, Supervision, Project administration, Methodology, Conceptualization

## Declaration of competing interest

The authors declare that they have no known competing financial interests or personal relationships that could have appeared to influence the work reported in this paper.

## Data availability

All the codes and data used to build the models are available on Electrifying Nigeria GitHub repository: https://github.com/stanage/ElectrifyingNigeria

## Declaration of generative AI and AI-assisted technologies in the writing process

During the preparation of this work the authors used Paperpal in order to improve the readability and language of the manuscript. After using this tool, the authors reviewed and edited the content as needed and take full responsibility for the content of the published article.

Table S11: Hyperparameters for SVR models of outputs without storage

| Output | Hyperparameter | Value |
|---|---|---|
| Total Annual Cost | C | 889908.6056884271 |
| | Epsilon | 0.11777148769815615 |
| | Kernel | rbf |
| | Degree | 1 |
| | Gamma | 0.017472079610730055 |
| $CO_2$ Emission | C | 567754.8073493874 |
| | Epsilon | 0.04917487660012416 |
| | Kernel | rbf |
| | Degree | 4 |
| | Gamma | 0.002912874794411995 |
| CEEP | C | 653379.5949488326 |
| | Epsilon | 0.006785960166006522 |
| | Kernel | rbf |
| | Degree | 1 |
| | Gamma | 0.008068725222563553 |
| Import | C | 831885.8768976415 |
| | Epsilon | 6.094759960876085 |
| | Kernel | rbf |
| | Degree | 1 |
| | Gamma | 0.005314081870230898 |

Table S12: Hyperparameters for XGB models of outputs without storage

| Output | Hyperparameter | Value |
|---|---|---|
| Total Annual Cost | Max depth | 3 |
| | Learning rate | 0.6559938747200145 |
| | Number of estimators | 742 |
| | Minimum child weight | 5 |
| | Gamma | 0.2364570538403641 |
| | Subsample | 0.9166005202188677 |
| | Column Sample by Tree | 0.5467984466797525 |
| | Alpha | 0.34210178318639684 |
| | Lambda | 0.6058891382532333 |
| | Random state | 590 |
| $CO_2$ Emission | Max depth | 8 |
| | Learning rate | 0.8058929837921041 |
| | Number of estimators | 707 |
| | Minimum child weight | 6 |
| | Gamma | 0.6203690244688376 |
| | Subsample | 0.8633675242642986 |
| | Column Sample by Tree | 0.351803542599373 |
| | Alpha | 0.38202482253134534 |
| | Lambda | 0.19326434501771286 |
| | Random state | 507 |
| CEEP | Max depth | 3 |
| | Learning rate | 0.5560764434875813 |
| | Number of estimators | 581 |
| | Minimum child weight | 1 |
| | Gamma | 0.17663683708839756 |
| | Subsample | 0.22400187438994715 |
| | Column Sample by Tree | 0.45123077147598545 |
| | Alpha | 0.8358287950607348 |
| | Lambda | 0.3030245049556971 |
| | Random state | 475 |
| Import | Max depth | 10 |
| | Learning rate | 0.43666973070690324 |
| | Number of estimators | 828 |
| | Minimum child weight | 8 |
| | Gamma | 0.28078926083310424 |
| | Subsample | 0.6681118230081214 |
| | Column Sample by Tree | 0.4876447096098242 |
| | Alpha | 0.29034765465801493 |
| | Lambda | 0.3669174369705459 |
| | Random state | 949 |

Table S13: Hyperparameters for MLP models of outputs without storage

| Output | Hyperparameter | Value |
|---|---|---|
| Total Annual Cost | Number of hidden layers | 4 |
| | Hidden layer 1 units | 8 |
| | Hidden layer 2 units | 30 |
| | Hidden layer 3 units | 50 |
| | Hidden layer 4 units | 32 |
| | Activation | relu |
| | Alpha | 0.13762346635059558 |
| | Solver | lbfgs |
| | Random state | 3 |
| $CO_2$ Emission | Number of hidden layers | 4 |
| | Hidden layer 1 units | 40 |
| | Hidden layer 2 units | 36 |
| | Hidden layer 3 units | 17 |
| | Hidden layer 4 units | 3 |
| | Activation | relu |
| | Alpha | 0.421732762333308 |
| | Solver | lbfgs |
| | Random state | 34 |
| CEEP | Number of hidden layers | 4 |
| | Hidden layer 1 units | 38 |
| | Hidden layer 2 units | 20 |
| | Hidden layer 3 units | 2 |
| | Hidden layer 4 units | 36 |
| | Activation | relu |
| | Alpha | 0.1368361449397597 |
| | Solver | lbfgs |
| | Random state | 100 |
| Import | Number of hidden layers | 4 |
| | Hidden layer 1 units | 33 |
| | Hidden layer 2 units | 10 |
| | Hidden layer 3 units | 8 |
| | Hidden layer 4 units | 40 |
| | Activation | relu |
| | Alpha | 0.20068666427656182 |
| | Solver | lbfgs |
| | Random state | 52 |

Table S14: Hyperparameters for SVR models of outputs with storage

| Output | Hyperparameter | Value |
|---|---|---|
| Total Annual Cost | C | 938266.6722019083 |
| | Epsilon | 7.634754757706834 |
| | Kernel | rbf |
| | Degree | 4 |
| | Gamma | 0.008893667810101987 |
| $CO_2$ Emission | C | 71459.27699166543 |
| | Epsilon | 0.9032951181705501 |
| | Kernel | rbf |
| | Degree | 1 |
| | Gamma | 0.0036156992519115353 |
| CEEP | C | 574688.30798438 |
| | Epsilon | 1.486938802734724 |
| | Kernel | rbf |
| | Degree | 4 |
| | Gamma | 0.008037353669894194 |
| Import | C | 996637.6632627285 |
| | Epsilon | 0.02434772338384239 |
| | Kernel | rbf |
| | Degree | 2 |
| | Gamma | 0.0014214789177860907 |

Table S15: Hyperparameters for XGB models of outputs with storage

| Output | Hyperparameter | Value |
|---|---|---|
| Total Annual Cost | Max depth | 3 |
| | Learning rate | 0.011723269811235474 |
| | Number of estimators | 875 |
| | Minimum child weight | 6 |
| | Gamma | 0.7712938662023604 |
| | Subsample | 0.7899217905693013 |
| | Column Sample by Tree | 0.8285058920843318 |
| | Alpha | 0.6370415102229366 |
| | Lambda | 0.3197877984587 |
| | Random state | 560 |
| $CO_2$ Emission | Max depth | 3 |
| | Learning rate | 0.7866691318932307 |
| | Number of estimators | 565 |
| | Minimum child weight | 7 |
| | Gamma | 0.9416813016248271 |
| | Subsample | 0.9492401329994024 |
| | Column Sample by Tree | 0.9677794380528386 |
| | Alpha | 0.18569042405228248 |
| | Lambda | 0.8638844510641983 |
| | Random state | 666 |
| CEEP | Max depth | 5 |
| | Learning rate | 0.21198957131622246 |
| | Number of estimators | 531 |
| | Minimum child weight | 1 |
| | Gamma | 0.83543860333434954 |
| | Subsample | 0.8560354221225375 |
| | Column Sample by Tree | 0.9500432957192226 |
| | Alpha | 0.0614153690574877 |
| | Lambda | 0.97283565837746876 |
| | Random state | 599 |
| Import | Max depth | 10 |
| | Learning rate | 0.31195099027474943 |
| | Number of estimators | 120 |
| | Minimum child weight | 8 |
| | Gamma | 0.5887647311991991 |
| | Subsample | 0.8053950657222284 |
| | Column Sample by Tree | 0.9531753908252204 |
| | Alpha | 0.9543398120967802 |
| | Lambda | 0.05317709105394063 |
| | Random state | 639 |

Table S16: Hyperparameters for MLP models of outputs with storage

| Output | Hyperparameter | Value |
|---|---|---|
| Total Annual Cost | Number of hidden layers | 4 |
| | Hidden layer 1 units | 5 |
| | Hidden layer 2 units | 5 |
| | Hidden layer 3 units | 22 |
| | Hidden layer 4 units | 46 |
| | Activation | relu |
| | Alpha | 0.6856023087910847 |
| | Solver | lbfgs |
| | Random state | 88 |
| $CO_2$ Emission | Number of hidden layers | 4 |
| | Hidden layer 1 units | 33 |
| | Hidden layer 2 units | 35 |
| | Hidden layer 3 units | 12 |
| | Hidden layer 4 units | 10 |
| | Activation | relu |
| | Alpha | 0.87732216286418 |
| | Solver | lbfgs |
| | Random state | 58 |
| CEEP | Number of hidden layers | 4 |
| | Hidden layer 1 units | 4 |
| | Hidden layer 2 units | 37 |
| | Hidden layer 3 units | 41 |
| | Hidden layer 4 units | 22 |
| | Activation | relu |
| | Alpha | 0.5394813016687909 |
| | Solver | lbfgs |
| | Random state | 75 |
| Import | Number of hidden layers | 4 |
| | Hidden layer 1 units | 7 |
| | Hidden layer 2 units | 28 |
| | Hidden layer 3 units | 23 |
| | Hidden layer 4 units | 19 |
| | Activation | relu |
| | Alpha | 0.9366596741565628 |
| | Solver | lbfgs |
| | Random state | 72 |